\documentclass{spie}

\usepackage{amsmath,amsfonts,amssymb}
\usepackage{graphicx}
\usepackage{url}
\usepackage[colorlinks=true, allcolors=blue]{hyperref}

\title{Progress on ELROI satellite license plate flight prototypes}

\author[a]{Rebecca M. Holmes}
\author[b]{Sawyer Gill}
\author[b]{James Z. Harris}
\author[c]{Joellen S. Lansford}
\author[b]{Riley Myers}
\author[a]{Charles T. Weaver}
\author[b]{Aaron P. Zucherman}
\author[b]{Anders M. Jorgensen}
\author[a]{David M. Palmer}
\affil[a]{Los Alamos National Laboratory, P.O. Box 1663, Los Alamos, NM 87545 USA}
\affil[b]{New Mexico Institute of Mining and Technology, Socorro, NM 87801 USA}
\affil[c]{New Mexico State University, Las Cruces, NM 88003 USA}

\authorinfo{Further author information: (Send correspondence to R.M.H.)\\
R.M.H.: E-mail: rmholmes@lanl.gov, Telephone: 1 505 667 3323\\ 
D.M.P.: E-mail: palmer@lanl.gov\\
A.M.J.: E-mail: anders.m.jorgensen@nmt.edu}

\begin{document} 
\maketitle

\begin{abstract}
The Extremely Low-Resource Optical Identifier (ELROI) beacon is a concept for a milliwatt optical ``license plate'' that can provide unique ID numbers for everything that goes into space. Using photon counting to enable extreme background rejection in real time, the ID number can be uniquely identified from the ground in a few minutes, even if the ground station detects only a few photons per second. The ELROI concept has been validated in long-range ground tests, and orbital prototypes are scheduled for launch in 2018 and beyond. We discuss the design and signal characteristics of these prototypes, including a PC-104 form factor unit which was integrated into a CubeSat and is currently scheduled to launch in May 2018, and basic requirements on ground stations for observing them. We encourage others to consider observing our test flights.
\end{abstract}


\section{INTRODUCTION}
\label{sec:intro}

   \begin{figure} [h]
   \begin{center}
   \begin{tabular}{c} 
   \includegraphics[width=0.7\textwidth]{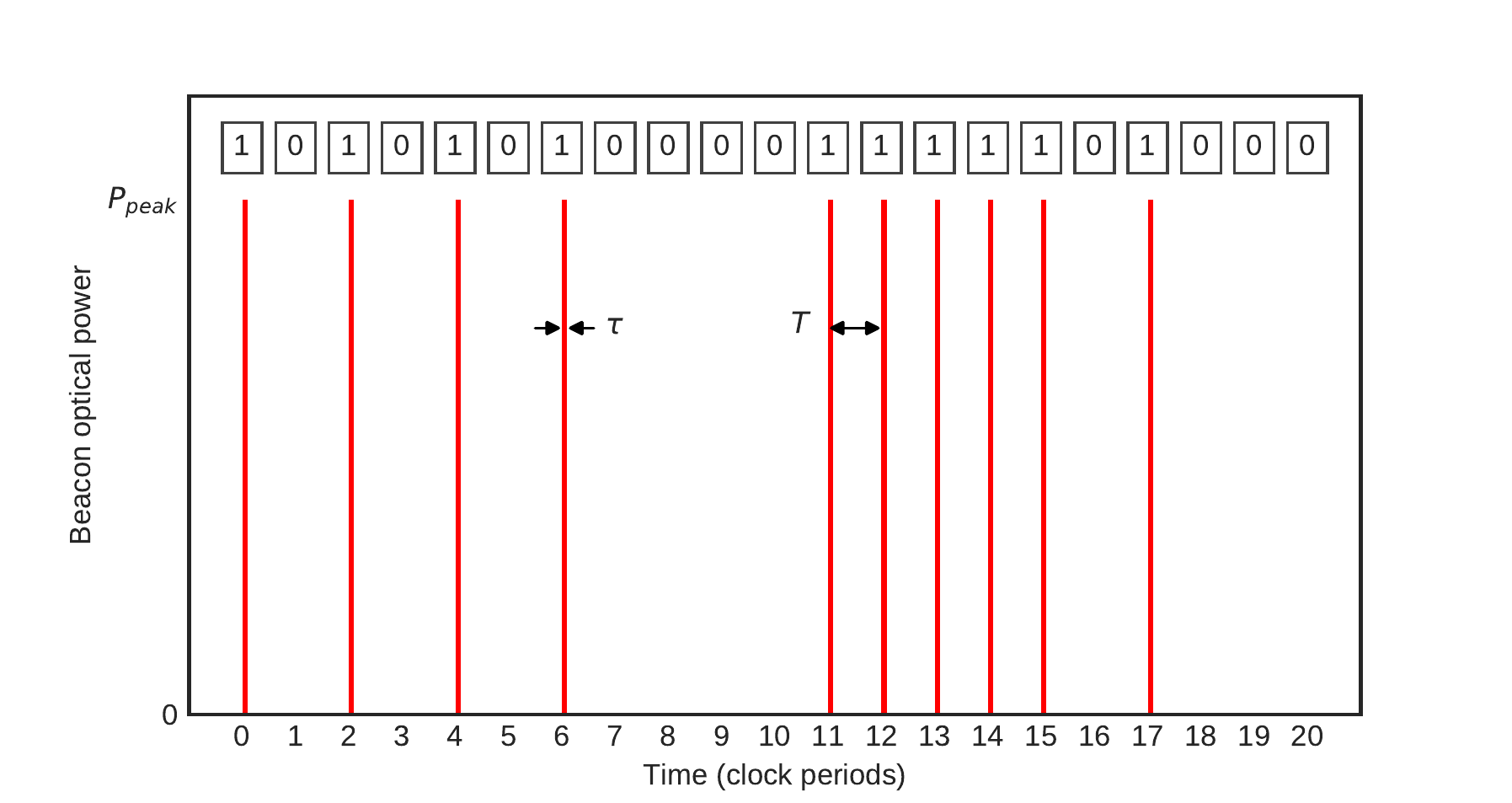}
   \end{tabular}
   \end{center}
   \caption[example] 
   { \label{fig:signal} 
An example of the signal produced by an ELROI beacon. The onboard laser diode emits short pulses of light (pulse width~=~$\tau$) separated by a fixed period (clock period~=~$T$). Each clock period encodes one bit of the beacon ID number: if the bit is a 1, the beacon emits a pulse, and if the bit is a 0, the beacon does not emit a pulse (the first 21 bits of a 128-bit ID are shown at top). The laser power during a pulse is $P_{\textrm{peak}}$; otherwise it is zero. The bit sequence (128 bits long for the current prototypes) repeats continuously several times per second. The pulse width in this example is exaggerated compared to the clock period. For a real beacon, a typical value of the ratio $\tau / T$ is 1:1000. The precise timing and high peak power of the signal pulses enable extreme background techniques that allow the ID number to recovered in single orbital pass even if only a few photons/second are detected at a ground station.}
   \end{figure} 

The Extremely Low-Resource Optical Identifier (ELROI) beacon is a milliwatt optical ``license plate'' that can provide unique ID numbers for everything that goes into space\cite{Palmer2015,Palmer2016,Palmer2017}. ELROI is designed to help address the problem of space object identification (SOI) in the crowded space around the Earth, where over 16,000 objects---from active satellites to rocket bodies and debris---are currently tracked and monitored\footnote{\url{https://www.space-track.org/}}. Tracking these objects requires continuous knowledge of each object's position and trajectory, and re-identifying a lost object is significantly easier if it carries an ID beacon that can be read from the ground. Small satellites such as CubeSats are also being launched in increasingly larger groups, and a typical CubeSat operator cannot identify their own satellite in the crowd without a beacon. There is currently no standard beacon technology that is small and light enough for the smallest satellites, and radio beacons have the additional drawback of RF interference.

ELROI is a concept for a tiny, autonomous optical beacon that uses short flashes of laser light to encode a unique ID number (Figure \ref{fig:signal}), which can be read from the ground by anyone with a small telescope and a photon-counting sensor. ELROI is smaller and lighter than a typical radio beacon, it is powered by its own small solar cell, and it can safely operate for the entire orbital lifetime of the host object. Using spectral filtering and photon counting to enable extreme background rejection in real time, the ID number can be uniquely identified in a few minutes, even if the ground station detects only a few photons per second. The ELROI concept has been validated in long-range ground tests, and orbital prototypes are scheduled for launch in 2018 and beyond.

A comprehensive overview of the ELROI encoding scheme and applications may be found in previously published references \cite{Palmer2017}. This paper will describe the design and signal characteristics of the current generation of ELROI prototypes, including a PC-104 form factor unit which was integrated into a CubeSat and is scheduled to launch no earlier than May 22, 2018, and an autonomous prototype, ELROI-UP, which can be attached to any host and does not require external power or control. We also discuss basic requirements on ground stations for observing ELROI, and we encourage others to consider observing our test flights.

\section{ELROI-PC104 on NMTSat}
\label{sec:pc104}

   \begin{figure} [ht]
   \begin{center}
   \begin{tabular}{c} 
   \includegraphics[height=5cm]{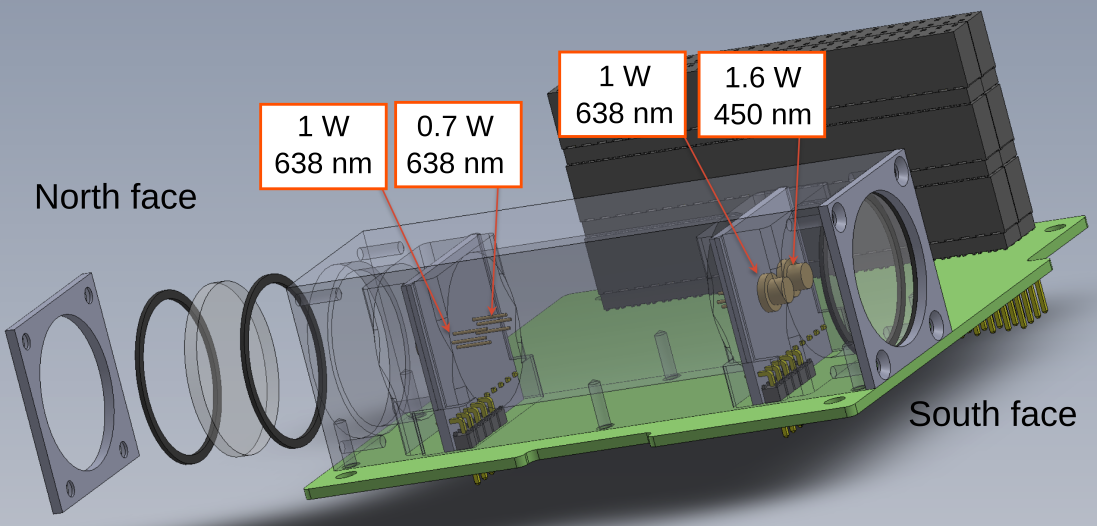}
   \end{tabular}
   \end{center}
   \caption[example] 
   { \label{fig:pc104} 
Components of the ELROI-PC104 board, with laser diodes shown.}
   \end{figure}
   
   \begin{figure} [ht]
   \begin{center}
   \begin{tabular}{c} 
   \includegraphics[width=0.5\textwidth]{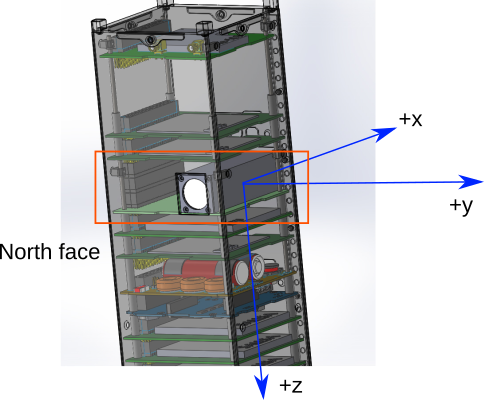}
   \end{tabular}
   \end{center}
   \caption[example] 
   { \label{fig:nmtsat} 
Partial schematic of NMTSat with local coordinate axes shown. The orange box indicates the position of the ELROI-PC104 board.}
   \end{figure}
   
ELROI-PC104 is a prototype designed in a standard PC-104 form factor, a common footprint for CubeSat payloads. Although the mature version of ELROI is designed to be autonomous and powered by its own small solar cell, this prototype receives power from the host satellite. It carries four laser diodes on two opposite faces (Figure \ref{fig:pc104} and Table \ref{diodes}). 

\begin{table}[hb]
\centering
\caption{Nominal wavelength ($\lambda$), peak power ($P_{\textrm{peak}}$), pulse width ($\tau$), clock period ($T$), and ID number for each laser diode on ELROI-PC104. The 128-bit ID numbers are transmitted with a binary on-off-key encoding (as depicted in Figure~\ref{fig:signal}), but are given in hexadecimal representation here for compactness. The three laser diodes with default state ``on'' will turn on after a 45-minute delay when ELROI-PC104 receives power. The other diode can be powered on by command from the host satellite.\newline}
\label{diodes}
\begin{tabular}{l|l|l|l|l|l|l}
ID number                									& $\lambda$ (nm) 	& $P_{\textrm{peak}}$ (W) & $\tau$ ($\mu$s) & $T$ ($\mu$s) & Position   	& Default 		\\ \hline
\texttt{e3386b221efa75323e0fc9aa5f28c2a4}  	& 638             			& 1.0       				& 	2 & 500	& North 		& On                 	\\
\texttt{1bf20e046fd9ee8154ad850bce95b6b2}   	& 638             			& 0.7       				& 	2 & 500 & North 		& On                 	\\
\texttt{a60de4513eb5de3a6ef6a056e2f08033}  	& 638             			& 1.0       				& 	2 & 500 & South 		& On                 	\\
\texttt{92d31ef48a08c4aedf2579a5d212dab9}		& 450             			& 1.6       				& 	2 & 500 & South 		& Off               
\end{tabular}
\end{table}

The ELROI-PC104 payload was built at Los Alamos National Laboratory and delivered in 2017 for integration into NMTSat, a 3U CubeSat designed and built by students at New Mexico Institute of Mining and Technology in Socorro, NM \cite{nmtsat}. NMTSat is funded by the NASA ELaNa CubeSat Launch initiative and is scheduled to launch on a Rocket Lab Electron mission no earlier than May 22, 2018. The primary purpose of NMTSat is educational, and in addition to ELROI, it carries payloads to monitor space weather and collect state-of-health data from the satellite (Figure \ref{fig:nmtsat}).
  
\subsection{Pre-launch optical measurements}

\begin{figure} [ht]
   \begin{center}
   \begin{tabular}{cc} 
   \includegraphics[width=0.4\textwidth]{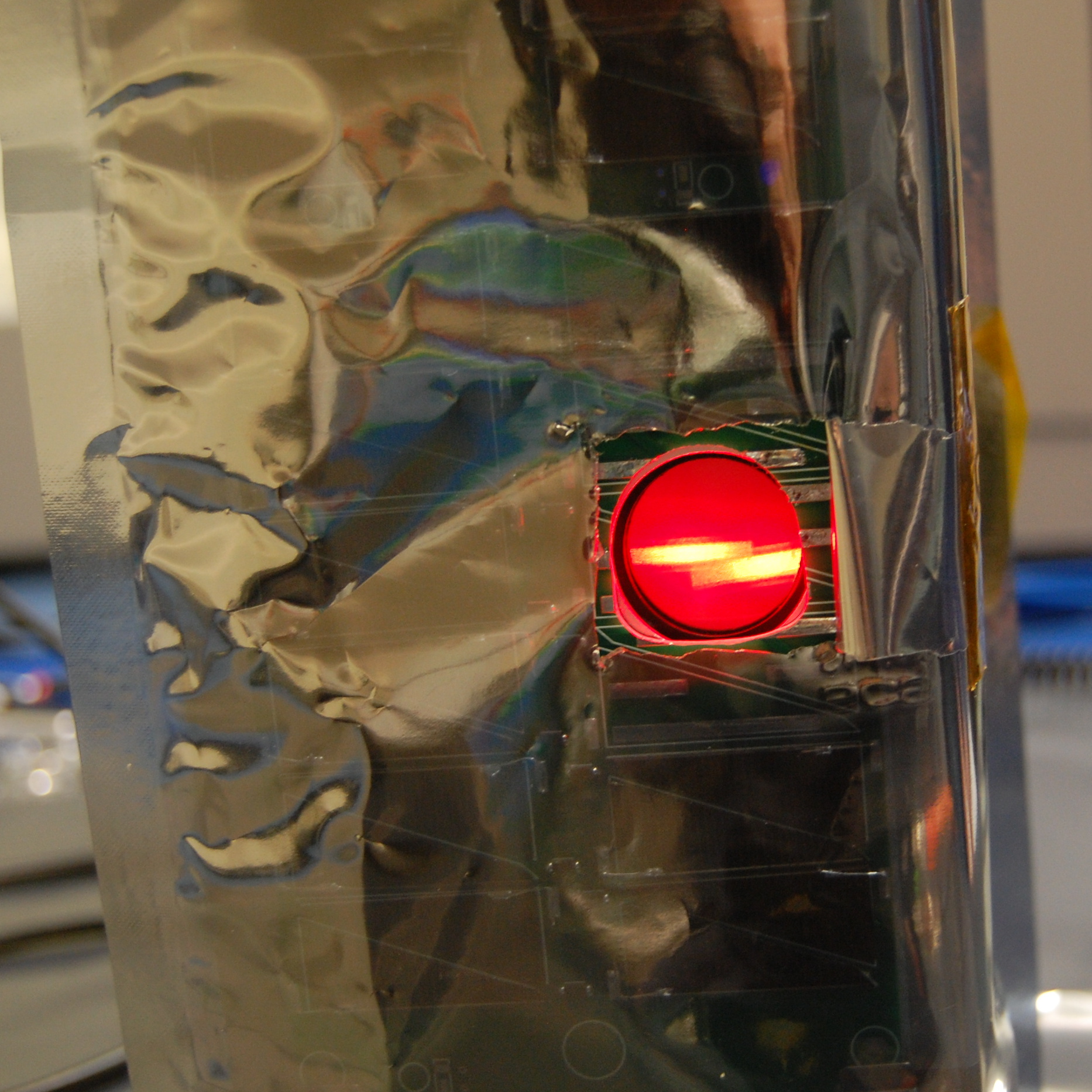} & \includegraphics[width=0.4\textwidth]{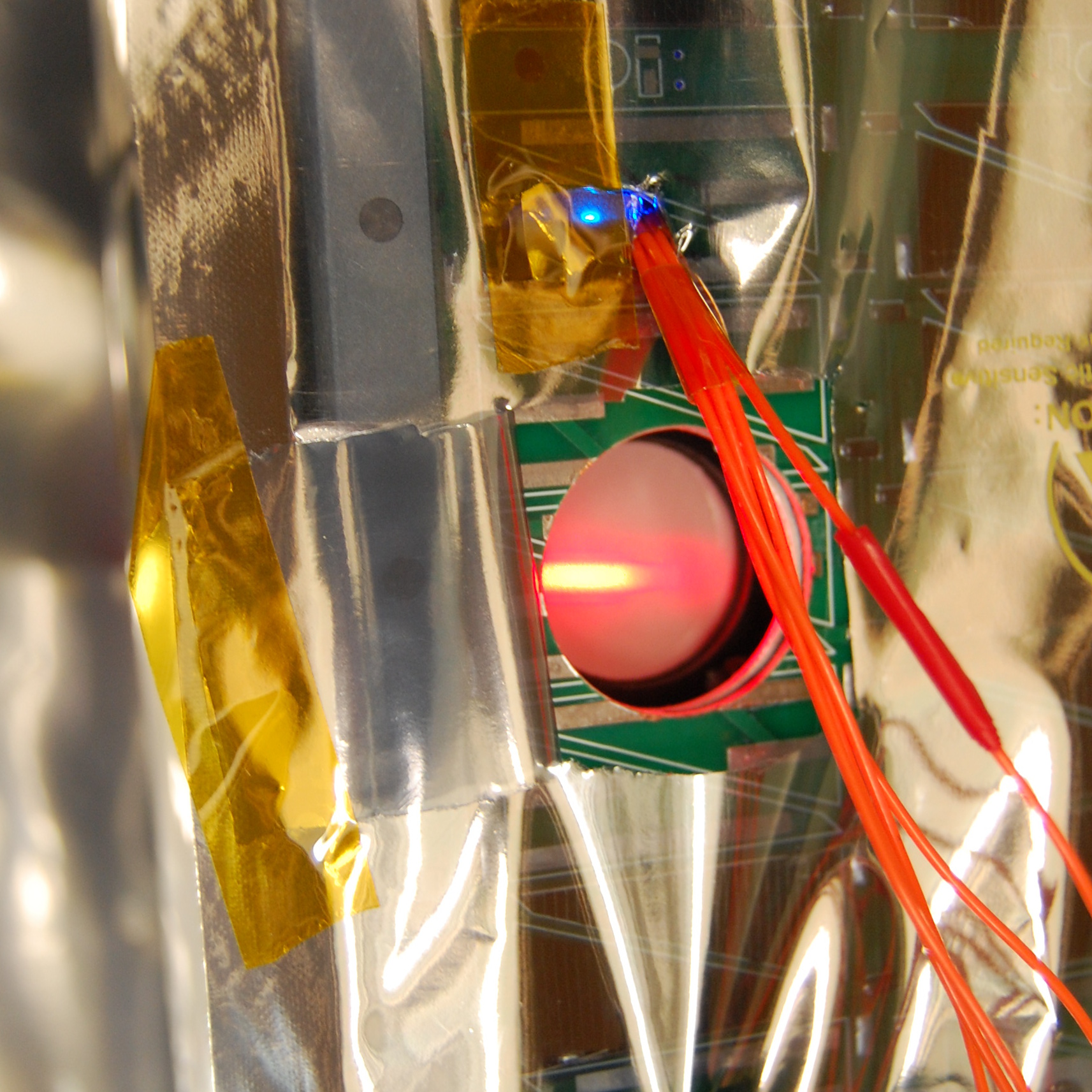}
   \end{tabular}
   \end{center}
   \caption[example] 
   { \label{fig:photos} 
ELROI laser diodes during pre-launch testing. The diode beams are visible on the diffusers. NMTSat is protected by anti-static material until delivery.}
   \end{figure}

   
To comply with the CubeSat standard, ELROI-PC104 is designed with a 45-minute delay between receiving power and turning on its laser diodes. After this delay, the laser diodes start up in a default configuration. ELROI-PC104 carries four laser diodes (Figure \ref{fig:pc104} and Table \ref{diodes}), but only the three 638-nm diodes are activated in the default state. The fourth, a 450-nm diode, draws more power and can be turned on by the NMTSat processor. ELROI-PC104 does not need to receive any commands from the host satellite in order to activate these three default laser diodes; if the payload receives power, they will turn on. There are two laser diodes behind each of two diffusers designed to scatter the laser light over a wide angle (Figure \ref{fig:photos}). Each laser diode is programmed to emit a different 128-bit binary ID number (given in Table \ref{diodes}).
   
\begin{figure} [ht]
   \begin{center}
   \begin{tabular}{cc} 
   \includegraphics[width=0.5\textwidth]{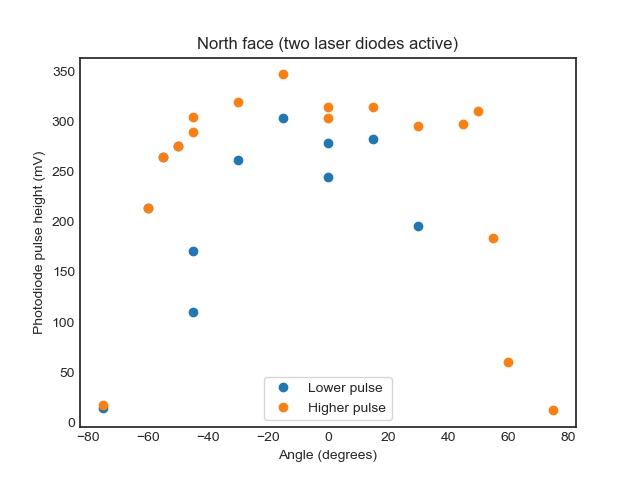} & \includegraphics[width=0.5\textwidth]{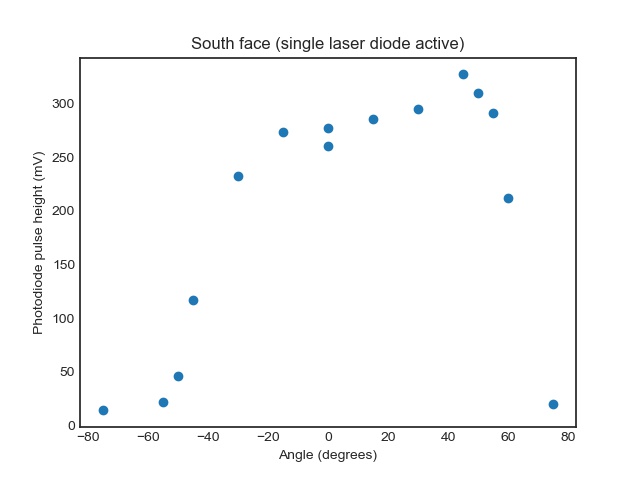}
   \end{tabular}
   \end{center}
   \caption[example] 
   { \label{fig:xyscans} 
Angular distributions of the light emitted from ELROI-PC104 in the xy plane. In the left figure, the ``higher pulse'' is presumed to correspond to the higher-power laser diode on that face (1 W peak power vs. 0.7 W). }
   \end{figure}

The angular distribution of the optical signal from each laser diode was measured in pre-launch testing, after integration into NMTSat. Engineered optical diffusers are used to spread the laser light over a wide angle, increasing the range of viewing angles over which it is visible from the ground. The results of these measurements are shown in Figure \ref{fig:xyscans}, and indicate that each laser diode emits over approximately $\pi$ steradians solid angle. The 450-nm laser diode on the south face of the satellite was not active during these tests.
   
\subsection{Polarity of the NMTSat attitude control magnet}

NMTSat contains a permanent magnet for simple attitude control using the Earth's magnetic field. The magnet is located near the top of the satellite as it is oriented in Figure \ref{fig:nmtsat}, and is co-aligned with the axis of the two ELROI-PC104 diffusers. The orientation of the poles of this magnet relative to the ELROI-PC104 diffusers determines which laser diodes will be visible from the ground at our observatory in New Mexico.

The north-seeking pole of the attitude control magnet is on the same side of NMTSat as the ELROI-UP diffuser with two 638-nm laser diodes (the north face of the satellite as shown in Figures \ref{fig:pc104} and \ref{fig:nmtsat}). This face is expected to be visible from the ground in the northern hemisphere if NMTSat reaches its predicted orbit (roughly circular at 500 km altitude and 83 degrees inclination). Therefore, we do not expect to be able to observe the 450-nm laser diode from New Mexico, but observers in the southern hemisphere may be able to see it.

\subsection{Planned observations}
\label{observations}

We will observe ELROI-PC104 from a Los Alamos National Laboratory ground station at Fenton Hill, near Jemez Springs, NM. Our receiver consists of a 36-cm aperture commercial telescope, optical bandpass filters, computerized mount, and a LANL-developed photon-counting camera \cite{Thompson2013}. (Equivalent cameras, while expensive, are available commercially\footnote{\url{https://www.photonis.com/en/product/imaging-photon-counting-camera}}.) As discussed in Section \ref{ground-station}, less expensive single- or few-element photon-counting sensors may also be used to observe ELROI, with correspondingly stricter requirements on the tracking and pointing accuracy of the telescope and mount system.

The optical link budget gives the expected photon detection rate under different conditions. The count rate in photons/second measured by the receiver is
\begin{equation}
R = P_{\textrm{avg}} \times 1/\Omega \times A/r^2 \times T_{\textrm{tot}} \times \varepsilon_{\textrm{DQE}}/E_{\gamma}
\end{equation}
$P_{\textrm{avg}}$ is the average power of the beacon, which depends on the peak power $P_{\textrm{peak}}$ and the duty cycle $d$. $\Omega$ is the solid angle of the beacon emission. $A$ is the collecting area of the receiver optics, and $r$ is the distance from the beacon to the receiver. $T_{\textrm{tot}}$ is the total optical transmission, including the spectral filter transmission $T_{\textrm{f}}$ and the atmosphere transmission $T_{\textrm{atm}}$. ($T_{\textrm{atm}}$ is ignored for the purposes of this link budget estimate, but it can be expected to contribute 1-3 dB of additional loss depending on conditions.) $\varepsilon_{\textrm{DQE}}$ is the quantum efficiency of the photon-counting detector, which ranges from 1\% (some PMTs) to 75\% (some SPADs), and has been measured to be 3.9\% for the LANL sensor. $E_{\gamma}$ is the energy per photon at the beacon wavelength.

Based on the values for ELROI-PC104 and NMTSat given in Table \ref{table:link-budget-parameters}, we expect to detect as many as 20 photons per second from ELROI-PC104 when it is directly over our ground station, which is sufficient to confidently recover an ID number with only 20 seconds or less of observation and provides margin for additional losses. Even if the expected signal count rate is reduced to 3 photons/second, the ID number can be recovered within 100-150 seconds, i.e., in a single pass over a ground station for a low Earth orbit (LEO) satellite \cite{Palmer2017}. 

\begin{table}[h]
\centering
\caption{Link budget parameters for ELROI-PC104 and NMTSat. The brighter of the two laser diodes on the north face of the satellite is used for $P_{\textrm{peak}}$. The distance to the receiver $r$ assumes a roughly circular orbit at 500 km altitude.}
\label{table:link-budget-parameters}
\begin{tabular}{llll}
\\ \hline
\textbf{Link budget parameter}               &                                 & \textbf{Value}       \\ \hline
Beacon wavelength                            & $\lambda$                       & 638 nm               \\
Peak power emitted                           & $P_{\textrm{peak}}$             & 1 W                  \\
Pulse width                                  & $\tau$                          & 2 $\mu$s             \\
Pulse interval                               & $T$                             & 500 $\mu$s           \\
Fraction of 1 bits                           & $f_1$                           & 0.50                 \\
Duty cycle                          & $d$                           & $f_1 \times \tau/T = 0.002$                \\
Solid angle of emission                      & $\Omega$                        & $\pi$ steradian               \\
Distance to receiver                         & $r$                             & 500-1000 km (above 30$^\circ$ elevation)             \\
Diameter of receiver telescope               & $D$                             & 36 cm                \\
Filter transmission at $\lambda$             & $T_{\textrm{f}}$                & 0.83                 \\
Filter bandwidth                             & $\Delta\lambda$                 & 10 nm                \\
Detector quantum efficiency                  & $\varepsilon_{\textrm{DQE}}$    & 0.039                
\end{tabular}
\end{table}

\subsection{Data Analysis}
The data from an observation consists of time-tagged photon detection events. Converting that data to an ID number requires several steps, all of which are efficient enough to do in real time:

\begin{enumerate}
  \item Detect and recover the beacon clock phase and period from the data.
  \item Using the measured clock phase and period, apply a phase cut to remove random background photons from the data (this reduces background by a factor $\tau/T$, typically $\sim$1/1000).
  \item Add up photon counts for each ID bit.
  \item Use photon counts to determine the most likely value (1 or 0) for each ID bit.
  \item Compare the recovered ID to a registry of known ELROI IDs, and determine which registry ID is the best match.
\end{enumerate}

A more detailed discussion of how to implement these steps is available in Reference [3]. The phase cut in Step 2, as well as the use of a narrowband optical filter to block background photons and transmit only the beacon laser wavelength, enables the extreme background rejection that makes ELROI possible.

The most significant source of background photons is reflected sunlight from the host satellite when it is illuminated. An exact prediction of the background rate from a given host is difficult, but for a CubeSat-sized host we expect to detect $\sim$100 background photons/second after a 10-nm bandpass filter with center wavelength 638~nm. Ground tests have demonstrated that the background rejection techniques discussed above can reduce this background rate to 0.4~photons/second. To provide tolerance for bit errors due to remaining background photons and/or the low rate of signal photons, an error-correcting code is used to generate the ID numbers. 



\section{ELROI-UP}
\label{sec:elroi-up}

   \begin{figure} [ht]
   \begin{center}
   \begin{tabular}{cc} 
   \includegraphics[width=0.4\textwidth]{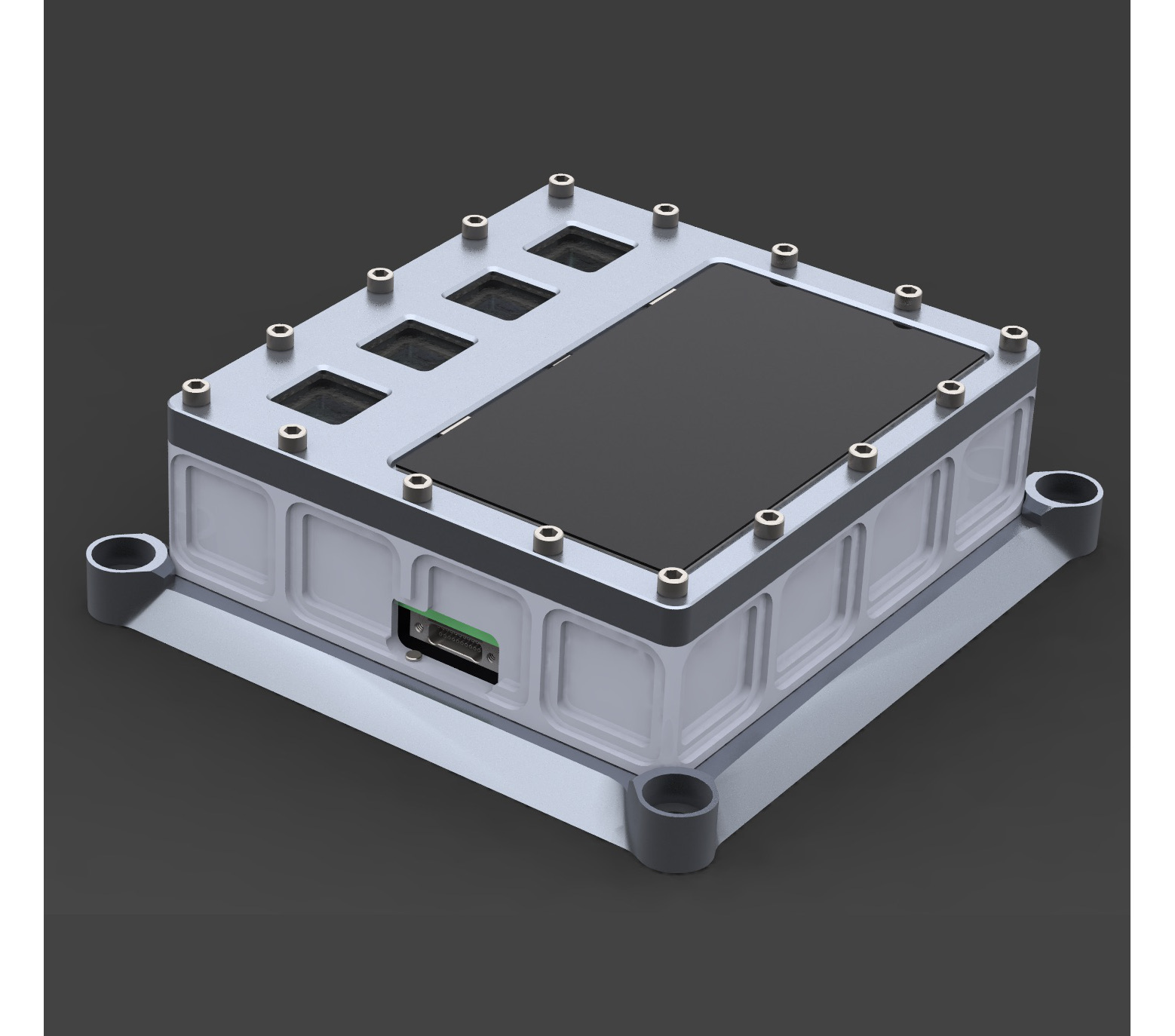} & \includegraphics[width=0.4\textwidth]{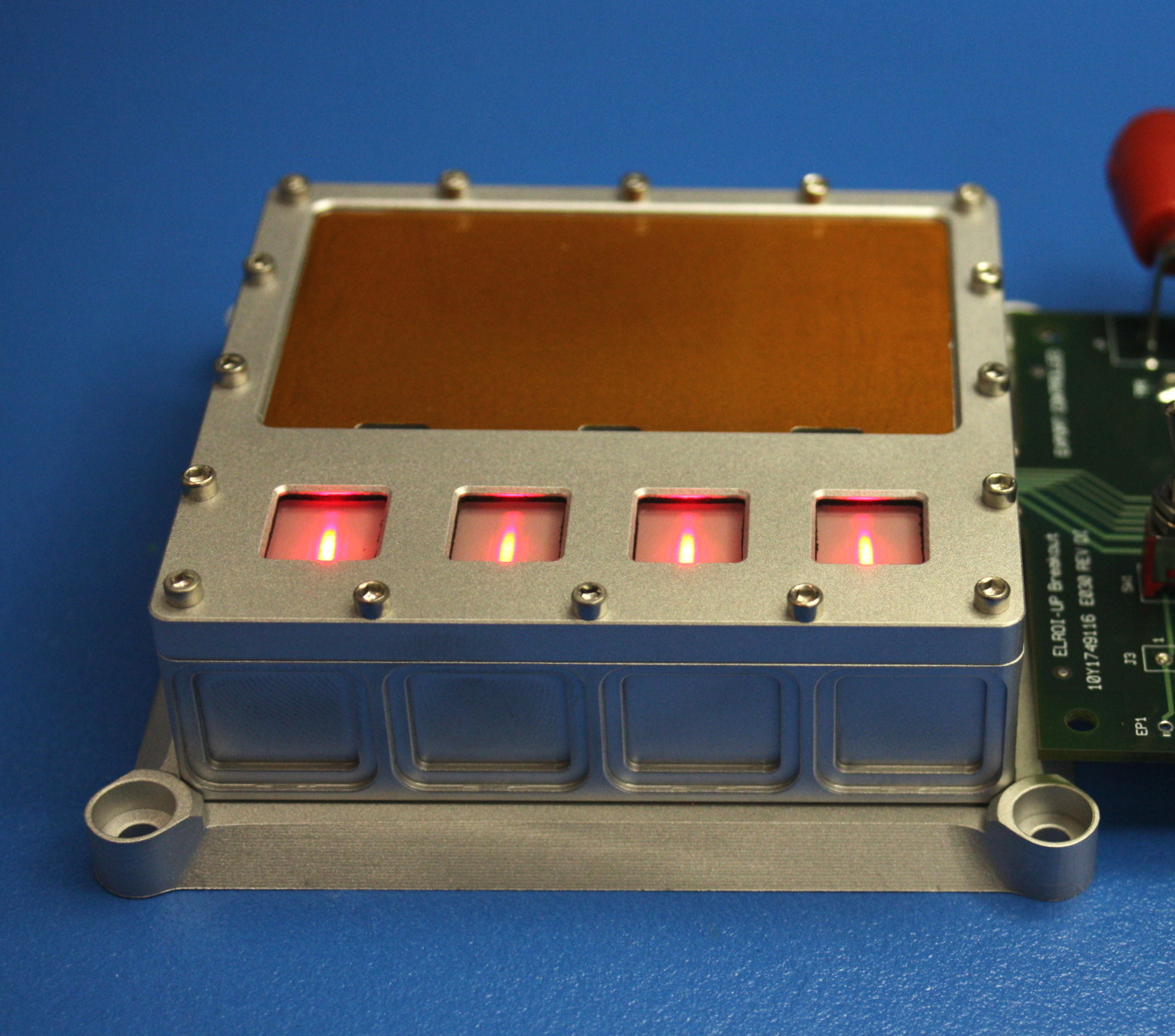}
   \end{tabular}
   \end{center}
   \caption[example] 
   { \label{fig:elroi-up} 
Left: Drawing of ELROI-UP. Right: Assembled ELROI-UP unit with laser diodes active. The unit shown here has not yet had its solar cell installed.}
   \end{figure}
   
\begin{figure} [ht]
   \begin{center}
   \begin{tabular}{c} 
   \includegraphics[width=0.6\textwidth]{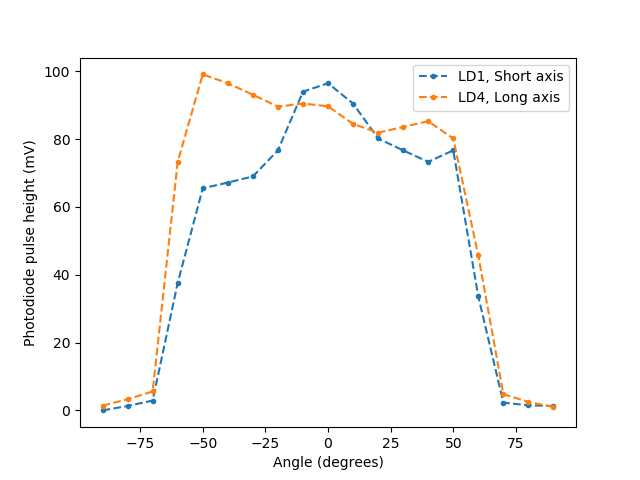}
   \end{tabular}
   \end{center}
   \caption[example] 
   { \label{fig:elroi-up-diff} 
Angular distributions of the light emitted from an ELROI-UP test unit. The two curves shown were measured from two of the four laser diodes. Short and long axis refers to the elliptical beam emitted by the diodes (visible in Figure \ref{fig:elroi-up}).}
   \end{figure}

The ELROI Universal Prototype (ELROI-UP) is a more advanced, fully autonomous design that is currently in final assembly and environmental testing (Figure \ref{fig:elroi-up}). ELROI-UP is equipped with its own small solar cell, and can receive power and commands from a host satellite, but does not require them. ELROI-UP can contain up to four laser diodes. The first test flight units will be populated with four $\lambda = $ 638 nm red laser diodes, at $P_{peak} = $~2.5~W. Different combinations of the four emitters will be pre-programmed to allow testing at up to $P_{peak} = $~10~W. The angular distribution of the optical signal was measured for a test unit (Figure \ref{fig:elroi-up-diff}). Each diode was measured to emit over approximately 1.3$\pi$ steradians solid angle.

ELROI-UP is $9.8\times9.2\times3.1 = 280 \textrm{ cm}^3$, the mass is $\sim$300 g, and the power (if externally supplied instead of provided by the solar cell) is 50 mW.  This SWaP (size, weight, and power) qualifies it as a true Low Resource Optical Identifier. After the prototype stage, the final target size of ELROI is a few centimeters square, similar to a thick postage stamp. The minimum size is limited by the solar cell needed to power the laser diode(s).

ELROI-UP is designed to be attached to any host that can accommodate it. The unit can also be mounted in a passive mechanical structure and launched as a free-flying CubeSat in 1/3U, 1/2U or larger form factor. We are willing to provide these units to interested launch opportunities.

\section{Summary of ELROI ground station requirements}
\label{ground-station}

As discussed in Section \ref{observations}, a large telescope is generally not required to read an ELROI ID. However, a larger ground station aperture does increase the signal count rate (as well as the background rate), and causes a corresponding decrease in the time required to read the beacon ID.

The sensitivity of the ELROI ID measurement relies on the use of a photon-counting detector, and the ELROI ID cannot be recovered with a conventional CCD camera (due to read noise).  Each photon registered by the detector must produce a discrete signal with time resolution better than the pulse width $\tau$. Our current ground station uses a LANL-developed photon-counting camera, which has a large-format sensor and good spatial resolution. Arrays of SPADs, position-sensitive readouts for PMTs, and other technologies can also provide spatial resolution. An imaging detector allows a less accurate tracking system, with the satellite photons separated from the sky background in post-processing. 

However, an imaging detector is not required to observe ELROI if the telescope tracking is sufficiently accurate to keep the host satellite within a small field of view for the duration of an observation. We expect single-pixel sensors to play a role in a future global ELROI system. Common single-pixel detectors include single-photon avalanche diodes (SPADs) and photomultiplier tubes (PMTs). Some suitable ground stations may already exist, needing only a small amount of extra hardware and software---for example, satellite laser ranging (SLR) stations have the equipment and expertise to track LEO satellites and observe them with single-pixel detectors \cite{Pearlman2002}.

We encourage others to consider observing ELROI-PC104 on NMTSat and any future ELROI flights, and to contact us to discuss possible ground station designs.

\acknowledgments 
 
Initial work on this project was supported by the US Department of Energy through the Los Alamos National Laboratory (LANL) Laboratory Directed Research and Development program as part of the IMPACT (Integrated Modeling of Perturbations in Atmospheres for Conjunction Tracking) project.  Further work was supported by the Richard P. Feynman Center for Innovation at LANL.  ELROI hardware and software was developed at LANL by Louis Borges, Richard Dutch, David Hemsing, Joellen Lansford and Charles Weaver, with thermal analysis by Alexandra Hickey, Lee Holguin, and Zachary Kennison.

\bibliography{ref} 
\bibliographystyle{hieeetr} 

\end{document}